\documentclass[a4paper,sort&compress]{aipproc}

\layoutstyle{6x9}

\usepackage{graphicx}
\usepackage{ifpdf}
\usepackage{subfigure}
\usepackage{url}

\ifpdf
	\usepackage{epstopdf}
\else
\fi

\newcommand{\wraprefprepost}[3]{\wrapprepost{#1}{#2}{\ref{#3}}}
\newcommand{\formatrefplain}{\ref}
\newcommand{\formatrefparens}{\wraprefprepost{(}{)}}

\newcommand{\wrapprepost}[3]{{#1}{#3}{#2}}

\newcommand{\tagwithlabel}[2]{#1~#2}

\newcommand{\makelabeledcrossrefmacro}[4]
	{\newcommand{#3}{#1{#4}{#2}}}
\newcommand{\makecrossrefmaker}[3]
	{\newcommand{#1}{\makelabeledcrossrefmacro{#2}{#3}}}

\newcommand{\eqnrefformat}{\formatrefparens}
\newcommand{\eqnlabelbinding}{\tagwithlabel}
\newcommand{\eqnlabel}{eq.}
\newcommand{\Eqnlabel}{Eq.}
\newcommand{\eqnslabel}{eqs.}
\newcommand{\Eqnslabel}{Eqs.}

\newcommand{\eqnnum}{\eqnrefformat}
\makecrossrefmaker{\newlabeledeqnref}{\eqnlabelbinding}{\eqnnum}
\makecrossrefmaker{\newwordpluseqnref}{\tagwithlabel}{\eqnnum}

\newlabeledeqnref{\eqn}{\eqnlabel}
\newlabeledeqnref{\Eqn}{\Eqnlabel}
\newlabeledeqnref{\eqns}{\eqnslabel}
\newlabeledeqnref{\Eqns}{\Eqnslabel}

\newwordpluseqnref{\andeqn}{and}
\newwordpluseqnref{\througheqn}{through}

\newcommand{\figrefformat}{\formatrefplain}
\newcommand{\figlabelbinding}{\tagwithlabel}
\newcommand{\figlabel}{fig.}
\newcommand{\Figlabel}{Fig.}
\newcommand{\figslabel}{figs.}
\newcommand{\Figslabel}{Figs.}

\newcommand{\fignum}{\figrefformat}
\makecrossrefmaker{\newlabeledfigref}{\figlabelbinding}{\fignum}
\makecrossrefmaker{\newwordplusfigref}{\tagwithlabel}{\fignum}

\newlabeledfigref{\fig}{\figlabel}
\newlabeledfigref{\Fig}{\Figlabel}
\newlabeledfigref{\figs}{\figslabel}
\newlabeledfigref{\Figs}{\Figslabel}

\newwordplusfigref{\andfig}{and}
\newwordplusfigref{\throughfig}{through}

\newcommand{\sxnrefformat}{\formatrefplain}
\newcommand{\sxnlabelbinding}{\tagwithlabel}
\newcommand{\sxnlabel}{section}
\newcommand{\Sxnlabel}{Section}
\newcommand{\sxnslabel}{sections}
\newcommand{\Sxnslabel}{Sections}

\newcommand{\sxnnum}{\sxnrefformat}
\makecrossrefmaker{\newlabeledsxnref}{\sxnlabelbinding}{\sxnnum}
\makecrossrefmaker{\newwordplussxnref}{\tagwithlabel}{\sxnnum}

\newlabeledsxnref{\sxn}{\sxnlabel}
\newlabeledsxnref{\Sxn}{\Sxnlabel}
\newlabeledsxnref{\sxns}{\sxnslabel}
\newlabeledsxnref{\Sxns}{\Sxnslabel}
	
\newwordplussxnref{\andsxn}{and}
\newwordplussxnref{\throughsxn}{through}

\newcommand{\tblrefformat}{\formatrefplain}
\newcommand{\tbllabelbinding}{\tagwithlabel}
\newcommand{\tbllabel}{table}
\newcommand{\Tbllabel}{Table}
\newcommand{\tblslabel}{tables}
\newcommand{\Tblslabel}{Tables}

\newcommand{\tblnum}{\tblrefformat}
\makecrossrefmaker{\newlabeledtblref}{\tbllabelbinding}{\tblnum}
\makecrossrefmaker{\newwordplustblref}{\tagwithlabel}{\tblnum}

\newlabeledtblref{\tbl}{\tbllabel}
\newlabeledtblref{\Tbl}{\Tbllabel}
\newlabeledtblref{\tbls}{\tblslabel}
\newlabeledtblref{\Tbls}{\Tblslabel}
	
\newwordplustblref{\andtbl}{and}
\newwordplustblref{\throughtbl}{through}

\renewcommand{\figlabel}{Fig.}
\renewcommand{\Figlabel}{Figure}
\renewcommand{\figslabel}{Figs.}
\renewcommand{\Figslabel}{Figures}

\usepackage{xspace}

\newcommand{\ie}{i.e.}
 
\newcommand{\eg}{e.g.}
\newcommand{\brim}{BRIM\xspace}
\newcommand{\RD}{R\&D\xspace}

\newcommand{\newterm}{\emph}
\newcommand{\subjectindex}[1]{\textsl{#1}}

\newcommand{\largeparens}[1]{\left ( #1 \right )}
\newcommand{\largesquare}[1]{\left [ #1 \right ]}

\newcommand{\largestraight}[1]{\left | #1 \right |}

\newcommand{\abs}{\largestraight}

\newcommand{\mathpuncspace}{\quad}
\newcommand{\mathcomma}{\mathpuncspace,}
\newcommand{\mathperiod}{\mathpuncspace.}

\newcommand{\mat}[1]{\mathbf{#1}}

\newcommand{\transpose}[1]{#1^\mathrm{T}}

\newcommand{\zeromat}{\mat{O}}

\newcommand{\vertexset}{\ensuremath{V}}
\newcommand{\edgeset}{\ensuremath{E}}
\newcommand{\graph}{\ensuremath{G}}
\newcommand{\vertexdegree}[1]{\ensuremath{d_{#1}}}

\newcommand{\adjmat}{\mat{A}}
\newcommand{\adjelem}[2]{A_{#1#2}}
\newcommand{\adjsubmat}{\mat{M}}

\newcommand{\probelem}[2]{P_{#1#2}}

\newcommand{\cardinality}{\largestraight}

\newcommand{\numedges}{\cardinality{\edgeset}}
\newcommand{\modularity}{\ensuremath{Q}}

\newcommand{\freq}[1]{f\largeparens{#1}}
\newcommand{\topfreq}[2]{\topfreqnoarg{#1}\largeparens{#2}}
\newcommand{\topfreqnoarg}[1]{f_{#1}}

\newcommand{\matrixbrackets}{\largesquare}

\newcommand{\bipartsubstructure}[1]{\matrixbrackets{
	\begin{array}{cc}
		\zeromat & #1 \\
		\transpose{#1} & \zeromat
	\end{array}}}

\newcommand{\topicmetric}[1]{d_{#1}}

\newtheorem{theorem}{Theorem}

\newtheorem{definition}[theorem]{Definition}

\graphicspath{{pdffigs/}{epsfigs/}{mpsfigs/}}

\begin{document}

\title{Searching for Communities in Bipartite Networks}


\author{Michael J. Barber}{
	address={Austrian Research Centers GmbH---ARC, Bereich systems research, Vienna, Austria}
}
\author{Margarida Faria}{
	address={CCM, Universidade da Madeira, Funchal, Portugal}
}
\author{Ludwig Streit}{
	address={CCM, Universidade da Madeira, Funchal, Portugal}
}
\author{Oleg Strogan}{
	address={CCM, Universidade da Madeira, Funchal, Portugal}
}

\date{\today}

\begin{abstract}
    Bipartite networks are a useful tool for representing and investigating interaction networks. We consider methods for identifying communities in bipartite networks. Intuitive notions of network community groups are made explicit using Newman's modularity measure. A specialized version of the modularity, adapted to be appropriate for bipartite networks, is presented; a corresponding algorithm is described for identifying community groups through maximizing this measure. The algorithm is applied to networks derived from the EU Framework Programs on Research and Technological Development. Community groups identified are compared using information-theoretic methods.
\end{abstract}

\maketitle


\section{Introduction}

Networks have attracted a burst of attention in the last decade (useful reviews include Refs.~\cite{ChrAlb:2007,DorMen:2004,New:2003,AlbBar:2002}), with applications to natural, social, and technological networks. Within biology, networks are prevalent, including: neural networks, where synapses link neurons; metabolic networks, describing metabolic processes in the cell, linking  chemical reactions and the regulatory processes that control them; protein interaction networks, representing physical interactions between an organism's proteins; transcription networks, describing regulatory interactions between different genes; food webs, using links to characterize who eats whom; and networks of sexual relations and infections, including AIDS models. Taking a broader view, networks seem to be everywhere! There are: electrical power grids, whose stability relates to the network structure; airline networks, with service efficiency tied to properties of the network; the World Wide Web, with search engines using the network links to locate pages; networks in linguistics, with words linked by co-occurrence; social networks of all sorts \citep{Bur:1992,Cou:2005,Gra:1973}; collaboration networks, describing joint works amongst actors, authors, research labs, \ldots; and many more. 

Of great current interest is the identification of community groups, or modules, within networks. 
Stated informally, a community group is a portion of the network whose members are more tightly linked to one another than to other members of the network. A variety of approaches \citep{AngBocMarPelStr:2007,GolKog:2006,Has:2006,NewLei:2007,ReiBor:2006,PalDerFarVic:2005,NewGir:2004,ClaNewMoo:2004,GirNew:2002} have been taken to explore this concept; see Refs.~\cite{DanDiaDucAre:2005,New:2004b} for useful reviews. Detecting community groups allows quantitative investigation of relevant subnetworks. Properties of the subnetworks may differ from the aggregate properties of the network as a whole, \eg,  modules in the World Wide Web are sets of topically related web pages.

Methods for identifying community groups can be specialized to distinct classes of networks, such as bipartite networks \cite{Bar:2007,GuiSalAma:2007}. The nodes in a bipartite network can be partitioned into two disjoint sets such that no two nodes within the same set are adjacent. Bipartite networks thus feature two distinct types of nodes, providing a natural representation for many affiliation or interaction networks, with one type of node representing actors and the other representing relations. Examples of actor-relation pairs include people attending events \citep{DavGarGar:1941,Fre:2003,DorBatFer:2004}, court justices making decisions \citep{DorBatFer:2004}, scientists jointly publishing articles \citep{New:2001a,New:2001}, organizations collaborating in projects \citep{BarKruKruRoe:2006,RoeBar:2007}, and legislators serving on committees \citep{PorMucNewFri:2007}. Arguably, bipartite networks are the empirically standard case for social networks and other interaction networks, with unipartite networks appearing---often implicitly---as projections.

\section{Community Structure in Networks}

\subsection{A Few Words on Graphs}

We formally describe networks using the language of graph theory. 
Let \( \vertexset \) be a set of \newterm{vertices} and  \( \edgeset \) be a set of vertex pairs or \newterm{edges} from \( \vertexset \times \vertexset \).  The pair \( \graph =\left(\vertexset, \edgeset\right) \) is called a \newterm{graph}.
In a \newterm{simple graph}, all pairs \( \left\{ u,v\right\} \in \edgeset \) are distinct and 
\( \left\{ u,u\right\} \notin \edgeset \), \ie, there are no double lines or loops.
Given a partition
\begin{equation}
	\vertexset=\vertexset_{1}+\vertexset_{2}
\end{equation}
where no edges exist between pairs of points within \( \vertexset_{i} \), \( i=1 \) or \( 2 \),
then \( \graph \) is said to be \newterm{bipartite}.

We shall consider simple graphs on a (large) finite set \( \vertexset \):
\begin{equation}
	\vertexset=\left\{ 1, 2, \ldots, n\right\}  \mathperiod
\end{equation}
The number of edges or \newterm{degree} \( \vertexdegree{v} \) of vertex \( v \) is defined by
\begin{equation}
	\vertexdegree{v}=\#\left( \left\{ v,\cdot \right\} \in E\right) 
\end{equation}
The number of edges is
\begin{equation}
	\left\vert \edgeset\right\vert =\frac{1}{2}\sum_{k=1}^{n}\vertexdegree{k} \mathperiod
\end{equation}
\( 2\left\vert \edgeset\right\vert  \) is often called the \newterm{volume} of the graph \( G \).

Graph structure is encoded in the \newterm{adjacency matrix}
\begin{equation}
	\adjmat=\left( \adjelem{i}{k}\right) _{1\leq i,k\leq n} \mathcomma
\end{equation}
with
\begin{equation}
	\adjelem{i}{k}=\left\{ 
	\begin{array}{ll}
	1 & \mbox{if }\left\{ i,k\right\} \in E \\ 
	0 & \mbox{otherwise}
	\end{array}
	\right . \mathperiod
\end{equation}
%

\subsection{Modularity}

Detecting community groups allows for the identification and quantitative investigation of relevant subnetworks. Local properties of the community groups may differ from the global properties of the complete network. For example, topically related web pages in the World Wide Web are typically interlinked, so that the contents of pages in distinct community groups  should reveal distinct themes. Thus, identification of community groups within a network is a first step towards understanding the heterogeneous substructures of the network.

To identify communities, we take as our starting point the \newterm{modularity}, introduced by \citet{NewGir:2004}. Modularity makes  intuitive notions of community groups precise by comparing network edges to those of a null model. As noted by  \citet{New:2006}:
\begin{quote}
	A good division of a network into communities is not merely one in which
	there are few edges between communities; it is one in which there are fewer
	than expected edges between communities. 
\end{quote}

\begin{definition}
The modularity~\( \modularity \) is---up to a normalization constant---the number of edges
within communities \( c \) minus those for a null model:
\begin{equation}
	\modularity\equiv \frac{1}{2\numedges }\sum_{c}\sum_{i,j\in c}\left(
	\adjelem{i}{j}-\probelem{i}{j}\right) \mathperiod
	\label{eq:modularity}
\end{equation}
\end{definition}

Along with \eqn{eq:modularity}, it is necessary to provide a null model, defining \( \probelem{i}{j} \). The standard choice for the null model constrains the degree distribution for the vertices to match the degree distribution in the actual network. Random graph models of this sort are obtained \citep{ChuLu:2002} by putting an edge between vertices \( i \) and \( j \) at random, with
the constraint that on average the degree of any vertex \( i \) is \( \vertexdegree{i} \). This constrains the expected adjacency matrix such that 
\begin{equation}
	\vertexdegree{i}=E\left( \sum_{j}\adjelem{i}{j}\right) \mathperiod
\end{equation}
Denote \( E\left( \adjelem{i}{j}\right)  \) by \( \probelem{i}{j} \) and assume further that \( P \)
factorizes into
\begin{equation}
	\probelem{i}{j}=p_{i}p_{j} \mathcomma
\end{equation}
leading to
\begin{equation}
	\probelem{i}{j}\equiv \frac{\vertexdegree{i}\vertexdegree{j}}{2\numedges } \mathperiod
\end{equation}
A consequence of the null model choice is that \( \modularity = 0 \) when all vertices are in the same community.

The goal now is to find a division of the vertices into communities such
that the modularity \( Q \) is maximal. An exhaustive search for a decomposition
is out of the question: even for moderately large graphs there are far too
many ways to decompose them into communities. Fast approximate algorithms do
exist (see, for example, Refs.~\citep{PujBejDel:2006,New:2004a}). 

\subsection{Finding Communities with \brim}

Specific classes of networks have additional constraints that can be reflected in the null model.
For bipartite graphs, the null model should be modified to reproduce the
characteristic form of bipartite adjacency matrices
\begin{equation}
	\adjmat=\bipartsubstructure{\adjsubmat}
	\mathperiod
	\label{eq:bipartsubstructure}
\end{equation}
Recently, specialized modularity measures and search algorithms have been proposed for
finding communities in bipartite networks \citep{Bar:2007,GuiSalAma:2007}. 

We make use of the algorithm called \brim: bipartite, recursively induced modules \citep{Bar:2007}. 
Starting from a (more or less) ad hoc partition of the vertices of type 1, it
is straightforward to optimize a corresponding decomposition of the vertices
of type 2. From there, optimize the decomposition of vertices of type 1, and
iterate. In this fashion, modularity  increases until a (local) maximum
is reached. However, the question remains: is the maximum a ``good'' one? At this level then a random search is called
for, varying the composition and number of communities, with
the goal of reaching a better maximum after a new round of hill
climbing using the \brim algorithm.

\section{An Initial Look at EU Research Networks}

In the ongoing research project NEMO \citep{nemo},
networks of research and development collaborations under EU Framework Programs FP1, FP2,\ldots, FP5 are studied. The collaborations in the Framework Programs give rise to bipartite graphs, with edges existing between projects and the organizations which take part in them. With this construction, participating
organizations are linked only through joint projects. 

In the various Framework Programs, the number of organizations 
ranges from 2,000 to 20,000, the number of projects ranges from 3,000 to 15,000, and the number of
links between them (project participations) ranges from 10,000 to 250,000 (see \tbl{tbl:fpstats} for precise values). A popular approach in social network analysis---where networks are often small, consisting of a few dozen nodes---is to visualize the networks and identify community groups by eye. However, the Framework Program networks are much larger: can we ``see'' the community groups in these networks?

\begin{table}
		\begin{tabular}{ccrrr}
			\hline
			\tablehead{1}{c}{c}{Framework \\Program} & \tablehead{1}{c}{c}{Period}     & \tablehead{1}{c}{c}{Projects} & \tablehead{1}{c}{c}{Organizations} \\ 
			\hline
			FP1                                & 1984--1987 & 3,283                     & 1,981                          \\ 
			FP2                                & 1987--1981 & 3,885                     & 4,572                          \\ 
			FP3                                & 1990--1994 & 5,529                     & 7,324                          \\ 
			FP4                                & 1994--1998 & 15,061                    & 19,755                         \\ 
			FP5                                & 1998--2002 & 15,559                    & 22,303                     \\
			\hline   
		\end{tabular}
		\caption{\RD projects and participants in the EU Framework Programs. The projects and organizations are used to define bipartite networks for each of the Framework Programs.}
		\label{tbl:fpstats}
\end{table}

Structural differences or similarities of such networks are not obvious at a
glance. For a graphical representation of the organizations and/or projects
by dots on an A4 sheet of paper, we would have to put these dots at a
distance of about \( 1\,\mathrm{mm} \) 
 from each other, and we then still would not have drawn the
links (collaborations) which connect them.

Previous studies used a list of coarse graining recipes to compact the networks into a form which would lend itself to a graphical representation \cite{BreCus:2004}.
As an alternative we have attempted to detect communities just using \brim,
\ie, purely on the basis of relational network structure, and blind with respect to
any additional information about the nature of agents.

\begin{figure}
	\includegraphics[width=\textwidth]{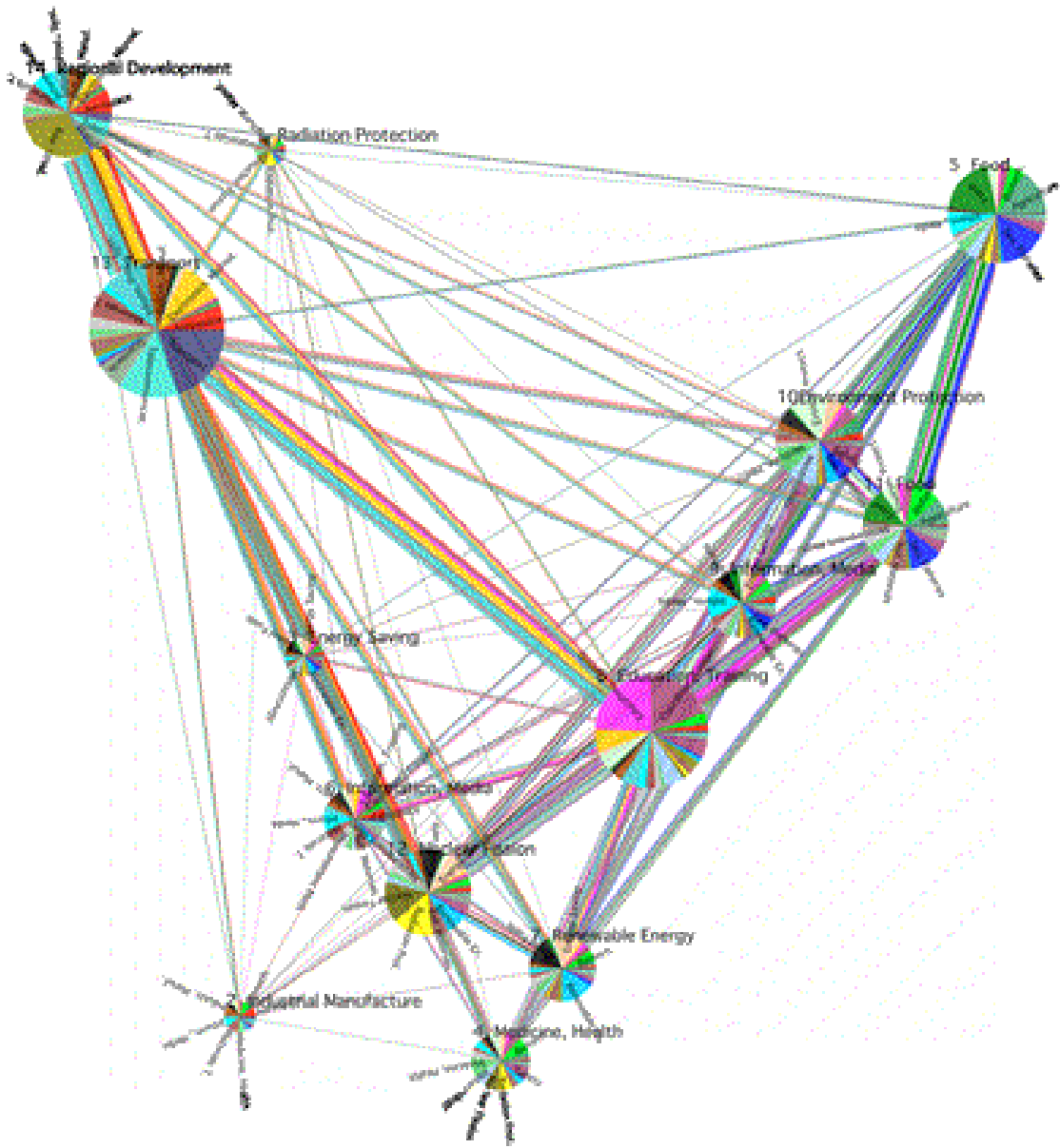}
	\caption{Community groups in the network of projects and organizations for FP3.}
	\label{fig:brimclusters}
\end{figure}

In \fig{fig:brimclusters}, we show a community structure for FP3 found using the \brim algorithm, with a modularity of \( \modularity = 0.602 \) for 14 community groups. The communities are shown as vertices in a network, with the vertex positions determined using spectral methods \citep{SeaRic:2003}.
The area of each vertex is proportional to the number of edges from the original network within the corresponding community. 
The width of each edge in the community network is proportional to the number of edges in the original network connecting community members from the two linked groups. 
The vertices and edges are shaded to provide additional information about their topical structure, as described in the next section. Each community is labeled with the most frequently occurring subject index.

\subsection{Topical Profiles of Communities}

Projects are assigned one or more standardized subject indices. There are 49 subject indices in total, 
ranging from \subjectindex{Aerospace} to \subjectindex{Waste Management}.

We denote by
\begin{equation}
	\freq{t} > 0 
\end{equation}
the frequency of occurrence of the subject index \( t \) in the network, with 
\begin{equation}
	\sum\limits_{t}\freq{t} = 1\mathperiod
\end{equation}
Similarly we consider the projects within one community \( c \) and the frequency 
\begin{equation}
	\topfreq{c}{t}\geq 0 
\end{equation}
of any subject index \(  t \) appearing in the projects only of that community. We
call \( \topfreqnoarg{c} \) the topical profile of community \( c \) to be compared with that of
the network as a whole. 

Topical differentiation of communities can be measured by comparing their
profiles, among each other or with respect to the overall network. This can be done
in a variety of ways \citep{GibSu:2002}, such as by the Kullback ``distance''
\begin{equation}
	D_{c}=\sum\limits_{t} \topfreq{c}{t}\ln \frac{\topfreq{c}{t}}{\freq{t}} \mathperiod
\end{equation}
A true metric is given by
\begin{equation}
	\topicmetric{c} = \sum\limits_{t} \abs{ \topfreq{c}{t}-\freq{t} } \mathcomma
\end{equation}
ranging from zero to two.

\begin{figure}
		\includegraphics[width=\textwidth]{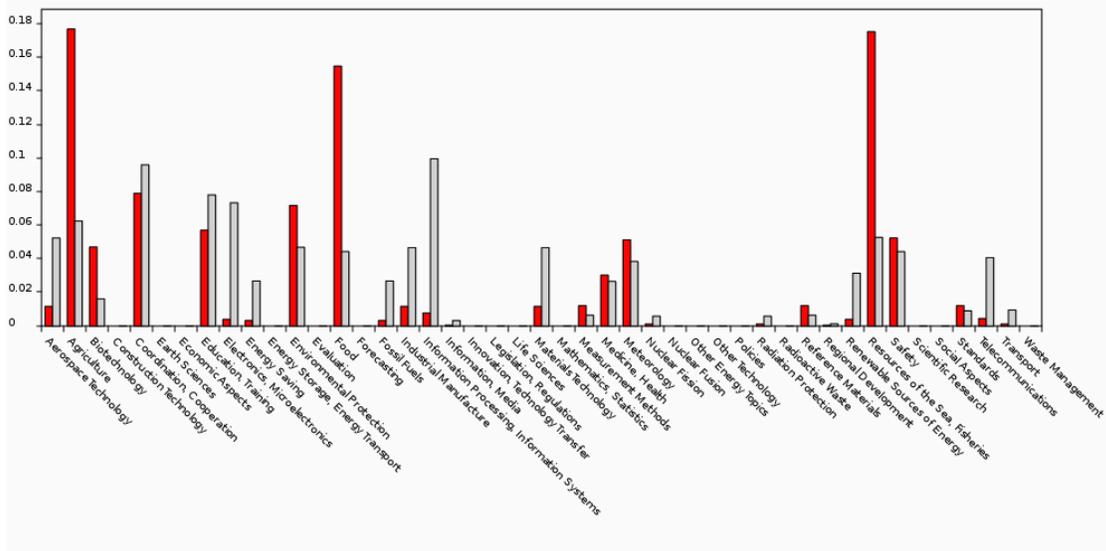}
	\caption{Topical differentiation in a network community. The histogram shows the difference between the topical profile \( \topfreq{c}{t} \) for a specific community (dark bars) and the overall profile \( \freq{t} \) for the network as a whole (light bars). The community-specific profile shown is for the community labeled ``11.~Food'' in \fig{fig:brimclusters}. The community has \( \topicmetric{c} = 0.90 \). }
	\label{fig:topichistogram}
\end{figure}

Topical differentiation is illustrated in \fig{fig:topichistogram}. In the figure, example profiles are shown, taken from the network in \fig{fig:brimclusters}. The community-specific profile corresponds to the community  labeled `11.~Food''  in \fig{fig:brimclusters}. Based on the most frequently occurring subject indices---\subjectindex{Agriculture}, \subjectindex{Food}, and \subjectindex{Resources of the Seas, Fisheries}---the community consists of projects and organizations focussed on \RD related to food products. The topical differentiation is \( \topicmetric{c} = 0.90 \) for the community shown.

\subsection{Projects and Subject Indices}

For further analysis, we have also looked for communities in 
networks of projects and subject indices. Here, the projects and subject indices constitute the vertices of a bipartite network, with edges existing between projects and the subject indices assigned to them. This construction disregards the organizations, providing an alternate approach to investigating the topical structure of the Framework Programs. 

\begin{figure}
	\includegraphics[width=\textwidth]{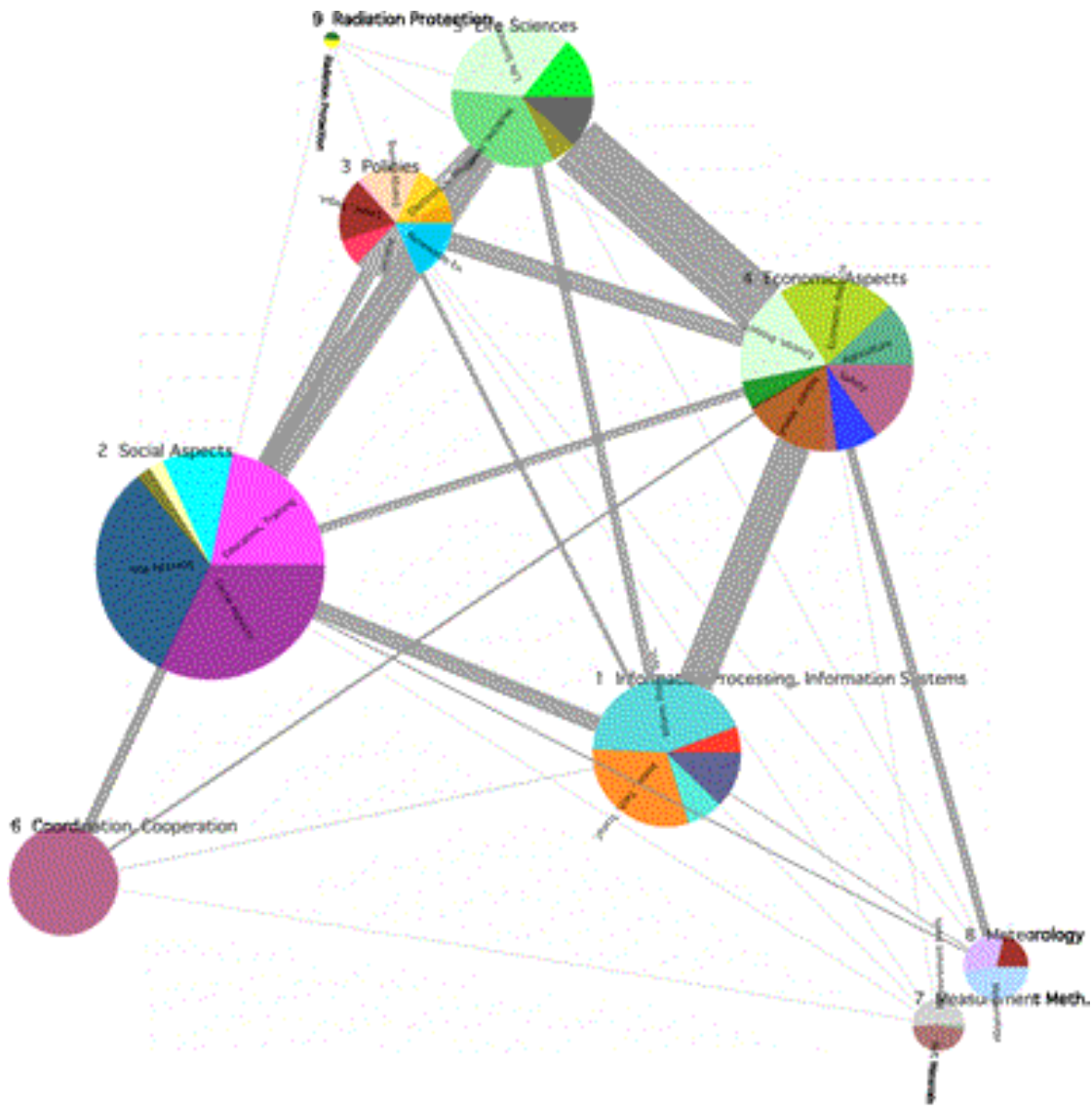}
	\caption{Community groups in the network of projects and subject indices for FP5.}
	\label{fig:si1}
\end{figure}

In \fig{fig:si1}, we show a set of nine community groups identified for the network of projects and subject indices in FP5. The modularity is \( \modularity = 0.50290 \) with these community groups. 

The later Framework Programs, such as FP5, show a fair degree of overlap between the communities, due to the subject indices being freely assigned in project applications. This is in marked contrast to the networks for the first three Framework Programs, where topics were attributed rigidly within thematic subprograms. The communities for FP1--3 thus have more clearly segregated community structures. FP1 is particularly extreme, having \emph{no} overlaps between the communities.  The differences between Framework Programs point to the need for some care in interpreting community structures: the communities in FP1--3 reflect policy structures while those found for later Framework Programs are more representative of interaction patterns.

%

\subsection{Comparing Decompositions with the Mutual Information}

For different decompositions of a (finite) set \( \Omega  \), a measure of
similarity is provided by the \newterm{normalized mutual information} \citep{DanDiaDucAre:2005}.
Set
\begin{equation}
	p(A)=\frac{\left\vert A\right\vert }{\left\vert \Omega \right\vert } \mathperiod
\end{equation}
For divisions of \( \Omega  \) into
\begin{enumerate}
\item J disjoint subsets \( X_{1},\ldots,X_{J} \) and
\item K disjoint subsets \( Y_{1},\ldots,Y_{K} \),
\end{enumerate}
compute the mutual information
\begin{equation}
	I_{0}(X,Y)=\sum\limits_{jk}p\left( X_{j}\cap Y_{k}\right) \log \frac{p\left(
	X_{j}\cap Y_{k}\right) }{p\left( X_{j}\right) p\left( Y_{k}\right) } 
	\label{eq:mutinfo}
\end{equation}
and normalize
\begin{equation}
	I(X,Y)=\frac{2I(X,Y)}{H(X)+H(Y)} \mathcomma
\end{equation}
where the entropy \( H \) is 
\begin{equation}
	H(X)=-\sum_{j}p\left( X_{j}\right) \log p\left( X_{j}\right) \mathperiod
	\label{eq:entropy}
\end{equation}
With the definitions in \eqns{eq:mutinfo} \througheqn{eq:entropy}, \( I \) ranges from zero, for uncorrelated decompositions of the set, to one, for perfectly correlated decompositions.

Using the \brim algorithm, we have partitioned network vertices into community groups by maximizing the modularity \( \modularity \). In principle, many dissimilar partitions of the vertices could produce similar modularity values. With the normalized mutual information (or a similar measure), we can assess the amount of shared structure between two different partitions of the vertex set.

\begin{figure}
	\includegraphics[width=\textwidth]{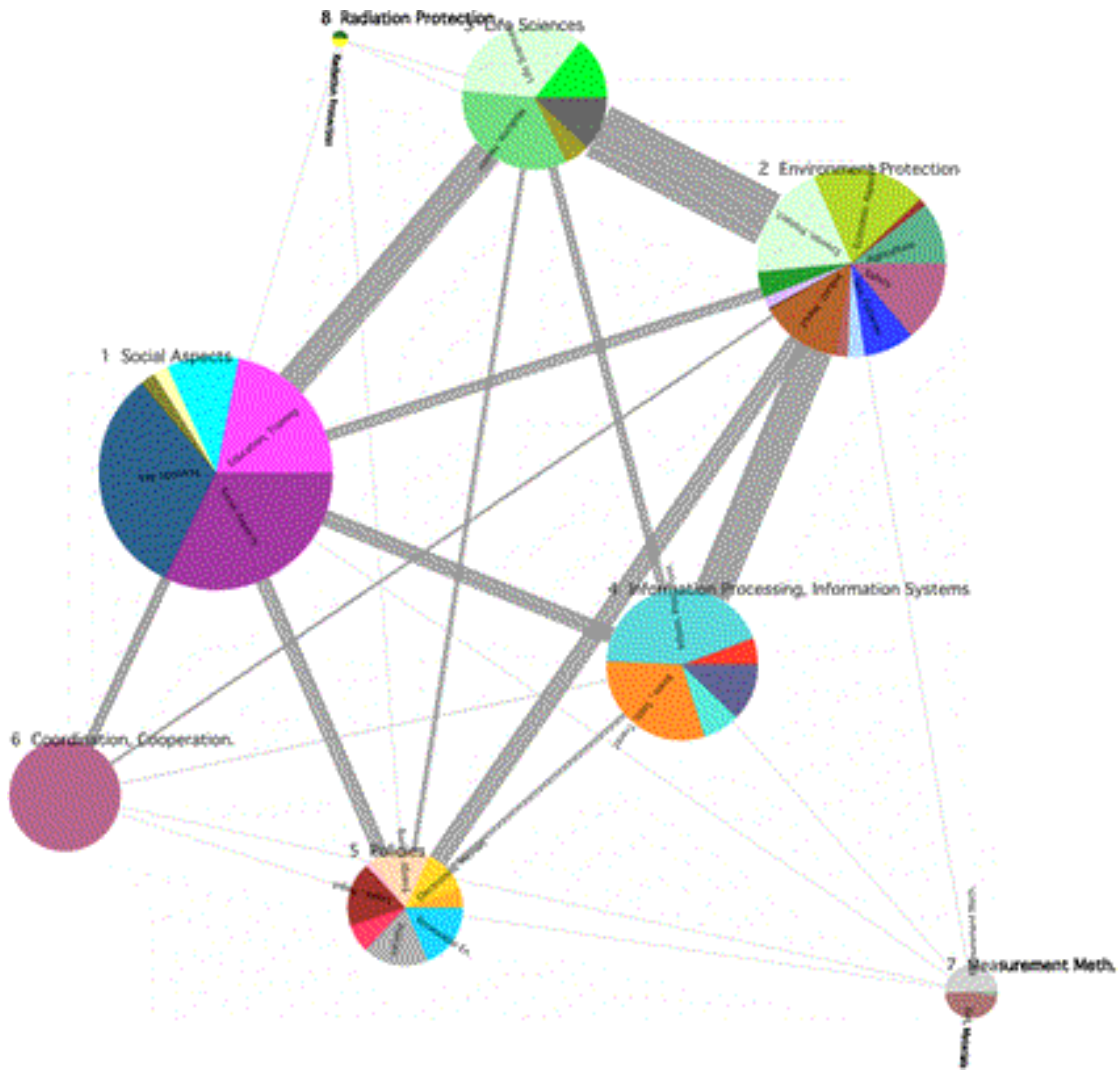}
	\caption{An alternate set of community groups in the network of projects and subject indices for FP5.}
	\label{fig:si2}
\end{figure}

For example, in \fig{fig:si1} we have shown a decomposition of the network of projects and subject indices for FP5. In \fig{fig:si2}, we show a second decomposition of the same network. The modularity is nearly identical---\( 0.50296 \) instead of the \( \modularity = 0.50290 \) seen previously---and the decompositions have some visible similarities. However, there are also definite structural differences, most prominently that the second decomposition has only eight communities while the first has nine. Are the structural differences significant? The normalized mutual information is found to be \( I = 0.98 \), indicating a strong correlation between the two decompositions and demonstrating that they have relatively minor structural differences.

\subsection{Outlook}

We have successfully identified community groups in networks defined from the Framework Program projects and the subject indices assigned to them. The full networks defined from the projects and the organizations taking part in them are considerably larger and correspondingly more challenging to investigate. For the full organizations-projects network, the \brim hill climbing algorithm is being supplemented with an aggressive, probabilistic search through community configurations upon which \brim acts. This extended search 
has only just begun, but preliminary results are encouraging.

Once established, communities will then be investigated with regard to their
internal structure, with the goal of identifying correlated properties within communities and contrasting properties across communities. We expect the analysis of internal structure to reveal patterns, themes, and motivations of collaborative research and development in the European Union.

\begin{theacknowledgments}
	The authors gratefully acknowledge financial support from the European FP6-NEST-Adventure Programme, under contract number~028875; from the Austrian Research Centers, in the \textit{Leitprojekt ``Bewertung und Gestaltung von Wissensnetzwerken''}; from the Portuguese FCT, under projects POCTI/MAT/58321/2004/FSE-FEDER and FCT/POCTI-219/FEDER; and from the Alexander von Humboldt Foundation, for travel support.
\end{theacknowledgments}


\bibliographystyle{aipproc}

\end{document}